\documentclass[12pt,3p,preprint,sort&compress]{elsarticle}
\usepackage[colorlinks=true]{hyperref}

\usepackage{amsmath}
\usepackage{amssymb}
\usepackage{mathtools}

\newcommand{\e}{{\mathrm{e}}}

\renewcommand{\d}{\partial}
\renewcommand{\l}{\left(}
\renewcommand{\r}{\right)}

\newcommand{\be}{\begin{equation}}
\newcommand{\ee}{\end{equation}}
\newcommand{\ba}{\begin{align}}
\newcommand{\ea}{\end{align}}
\newcommand{\bg}{\begin{gather}}
\newcommand{\eg}{\end{gather}}
\newcommand{\bseq}{\begin{subequations}}
\newcommand{\eseq}{\end{subequations}}

\renewcommand{\th}{\mathop{\mathrm{tanh}}\nolimits}
\newcommand{\ch}{\mathop{\mathrm{cosh}}\nolimits}
\newcommand{\sh}{\mathop{\mathrm{sinh}}\nolimits}

\newcommand{\half}{\frac{1}{2}}

\newcommand{\cH}{{\mathcal{H}}}
\newcommand{\bx}{{\mathbf{x}}}
\newcommand{\bk}{{\mathbf{k}}}

\begin{document}

\title{Some like it hot: $R^2$ heals Higgs inflation, but does not cool it}

\author[uom]{Fedor Bezrukov}
\ead{Fedor.Bezrukov@manchester.ac.uk}
\author[inr,mpti]{Dmitry Gorbunov}
\ead{gorby@ms2.inr.ac.ru}
\author[uom]{Chris Shepherd}
\ead{christopher.shepherd-3@postgrad.manchester.ac.uk}
\author[inr]{Anna Tokareva}
\ead{tokareva@ms2.inr.ac.ru}
\address[uom]{The University of Manchester, School of Physics and Astronomy, Manchester M13 9PL, United Kingdom}
\address[inr]{Institute for Nuclear Research of Russian Academy of Sciences, 117312 Moscow, Russia}
\address[mpti]{Moscow Institute of Physics and Technology, 141700 Dolgoprudny, Russia}

\begin{abstract}
  Strong coupling in Higgs inflation at high energies hinders a joint description of inflation, reheating and low-energy dynamics. The situation may be improved with a proper UV completion of the model. A well-defined self-consistent way is to introduce an $R^2$-term into the
  action.  In this modified model the strong coupling scale returns
  back to the Planck scale, which justifies the use of the
  perturbative methods in studies of the model dynamics after
  inflation.  We investigate the reheating of the post-inflationary
  Universe, which involves two highly anharmonic oscillators strongly
  interacting with each other: homogeneous Higgs field and
  scalaron. We observe that in interesting regions of model
  parameter space these oscillations make longitudinal components of
  the weak gauge bosons tachyonic, triggering instant preheating at
  timescales much shorter than the Hubble time.  The weak gauge
  bosons are heavy and decay promptly into light Standard Model
  particles, ensuring the onset of the radiation domination era right after
  inflation.
\end{abstract}

\maketitle

\section{Introduction}
\label{Sec:Introduction}

A large variety of inflationary models lead to dynamics which make
the Universe spatially flat and homogeneous, simultaneously producing
matter and gravitational perturbations consistent with present
observations of the cosmic microwave background and large scale structure
\cite{Akrami:2018odb,Alam:2016hwk,Ross:2014qpa}.  Naturally, viable
inflationary models with additional signatures deserve special
attention. In particular, independent tests of the inflationary
dynamics are provided in models with (a part of the) inflaton sector
playing some role in late-time cosmology or low-energy particle
phenomenology, see e.g.~\cite{Murayama:1993xu,Bezrukov:2009yw,Lerner:2009xg,Tenkanen:2016twd,Takahashi:2019qmh}. 
Among these models, Higgs-driven inflation
\cite{Bezrukov:2007ep} is in a unique position since the main
component in this inflationary model has been discovered
\cite{Aad:2012tfa,Chatrchyan:2012xdj} and its properties were
extensively studied at LHC \cite{Tanabashi:2018oca}.

While very appealing indeed, the concept of
Higgs inflation has unresolved issues both on phenomenological and
theoretical sides. The phenomenological tension arises due to the
Standard Model (SM) parameters -- the Higgs boson mass, top quark
mass, strong and weak gauge couplings -- whose central values measured
at the electroweak scale imply that quantum corrections to the Higgs
boson self-coupling make it negative for large values of the Higgs
field \cite{Krasnikov:1978pu,Hung:1979dn,Politzer:1978ic,Bednyakov:2015sca}.  Thus, its energy
density becomes negative, causing problems for cosmology in general \cite{Herranen:2015ima} and for implementation of plain
Higgs inflation. 

The theoretical problems are associated with the large dimensionless
constant of the non-minimal coupling to gravity. First, the large
coupling constant makes the theory strongly coupled much below the
Planck scale \cite{Burgess:2009ea,Barbon:2009ya}. Second, the model is
non-renormalizable because of the non-minimal coupling to gravity,
which decouples the parameters of the small and large field potentials
at the quantum level. Hence, the cosmological observables (amplitudes and tilts of the perturbation spectra) and low-energy observables to be measured in particle physics experiments are not related to each other, which prevents direct tests of the model. 

While one may argue that the true values of the relevant physical
parameters are actually at about $2\sigma$ off their presently
accepted central values, making the Higgs potential positive all the way
to the Planck scale, the large coupling constant is the key ingredient
of the model and cannot be changed at will. However, it was argued
\cite{Bezrukov:2010jz} that the strong coupling scale grows with the
Higgs field value.  Recall that the first investigations of the
post-inflationary dynamics and reheating of the Universe
\cite{Bezrukov:2008ut,GarciaBellido:2008ab} yield a consistent
estimate of the reheating temperature, which demonstrates that from
the inflation till the radiation domination stage the system
dynamics never exhibit the strong coupling behaviour. It was also
shown \cite{Bezrukov:2011sz} that the presence of non-renormalizable
operators suppressed by the field-dependent scale has no impact on the
inflationary predictions for the spectral parameters of scalar and tensor
perturbations.

However, the first estimates of the reheating in
Refs.~\cite{Bezrukov:2008ut,GarciaBellido:2008ab} were incomplete.
The crucial ingredient -- the evolution of longitudinal components of the
weak gauge bosons -- was missed there. The dependence of this 
component on the external Higgs field (inflaton) turns
out to be spiky
\cite{Ema:2016dny,DeCross:2016cbs,Sfakianakis:2018lzf}, leading to
violent production of longitudinal modes of the weak gauge bosons and
very rapid reheating of the Universe. This finding is dangerous due to
the strong coupling problem, since the reheating dynamics happen
right inside the strong coupling domain, making any analysis
unreliable. At the same time, the cosmological predictions of an
inflationary model depend on the evolution of the Universe after inflation
and the reheating temperature, so knowledge of the proper reheating
dynamics is essential for precise calculation of the inflationary
observables.

All these problems ask for a modification of Higgs inflation
capable of keeping the model inside the weak coupling regime from
inflation, through preheating, till the onset of the hot stage.  Such
a modification has been recently suggested
\cite{Ema:2017rqn,Gorbunov:2018llf} with a key ingredient -- an
$R^2$ term\footnote{A similar
  construction can be achieved with an additional scalar field
  \cite{Giudice:2010ka}, leaving more free parameters in the model.} added to the model Lagrangian.\footnote{As a bonus, the model addresses the phenomenological issue of the negative Higgs coupling too, somewhat reducing the aforementioned 2$\sigma$-tension \cite{Gorbunov:2018llf}.}
Detailed studies of the inflationary evolution and perturbative unitarity in the
scalar, gravity and gauge sectors revealed a region in the model
parameter space \cite{Gorbunov:2018llf} where the model remains weakly
coupled up to the Planck scale and its inflationary dynamics are similar
to the original Higgs inflation.  These conditions open a possibility
to test this cosmological model directly in particle physics
experiments.

Here we refine the predictions of Higgs inflation UV-completed by the $R^2$-term.  Namely, we study the post-inflationary dynamics of the model and reheating of the Universe.
In the interesting range of the model parameter space we observe exponentially fast generation of radiation \emph{due to the tachyonic instability} of the longitudinal components of the weak bosons\footnote{ The Higgs bosons are also produced in the described tachyonic regime, but as a subdominant component.} and their subsequent decays into light SM particles.

For particular dynamics at the moment of the scalaron crossing zero, the tachyonic instability is sufficiently drastic to complete preheating during less than one period of the scalaron oscillation. These special dynamics occur in the vicinities of the bifurcation points between two branches of the solutions, and are realised at the first scalaron crossing for particular ranges of the theory parameters. In general, this situation may not be realised at the first scalaron crossing, but preliminary calculations suggest that it should generally occur after only a few scalaron oscillations. While the exact number of oscillations cannot be known for given model parameters without accounting for backreaction, preliminary estimations suggest that reheating generally takes less than one Hubble time. This statement is more certain in the Higgs-like limit \cite{Ema:2017rqn,Gorbunov:2018llf}, and means that reheating in this model is instantaneous from a cosmological viewpoint.	Therefore, even though both $R^2$-healed and ``pure'' Higgs inflation reheat via the generation of longitudinal weak gauge bosons, the underlying dynamics are rather different.
In particular, we confirm that the spike in the mass of the longitudinal gauge bosons \cite{Ema:2016dny,DeCross:2016cbs,Sfakianakis:2018lzf} does not produce the weak bosons in the amount sufficient for reheating \cite{He:2018mgb}.

We emphasize that the current description does not take into account the details of the backreaction of the produced particles on the background.  In particular, it is assumed that, as far as the production of the weak bosons is sourced by the Higgs field at short timescales, there is no immediate energy transfer from the (slower) oscillations of the scalaron background.  First, this observation by itself requires that reheating happens during several scalaron oscillations, unless the model parameters correspond to a very close vicinity of the tachyonic bifurcation at the first scalaron zero crossing.  Second, even in the most optimistic case the backreaction must be taken into account because the intensive energy flow from the homogeneous mode to particles changes the scalar field dynamics and consequently the production process itself.  Also, there is additional energy drain from the Higgs oscillations away from scalaron zero crossing due to parametric resonance, which is subdominant for direct reheating process, but changes the details of background homogeneous field evolution. All this issues require detailed investigation, which we leave for the further study. However, given the dramatic swiftness of the process we expect the instant preheating of the Universe right after inflation to be the general model prediction in the Higgs-like region of the model parameters.

The paper is organised as follows. In Sec.~\ref{Sec:II} we describe
the model and recall its inflationary dynamics. We present the
equations of motion for homogeneous scalar fields and reduce them for
the small field limit suitable for investigation of the background
evolution after inflation. Sec.~\ref{Sec:III} contains equations for
inhomogeneous modes in the scalar and gauge boson sectors. Sec.~\ref{Sec:IV} describes the particle production. The obtained numerical results are discussed in Sec.~\ref{Sec:V}. We conclude and finally present the predictions for the cosmological parameters in Sec.~\ref{Sec:Conclusion}.


\section{Model Lagrangian and evolution of scalar homogeneous modes}
\label{Sec:II}

Higgs inflation, augmented with a term quadratic in curvature (Ricci)
scalar $R$, is described by the action
\be
  \label{action-1}
  S_0=\int d^4x \sqrt{-g} \l-\frac{M_P^2+\xi h^2}{2}R+\frac{\beta}{4}
  R^2+\frac{1}{2}g^{\mu\nu}\d_{\mu}h\d_\nu h-\frac{\lambda}{4}h^4\r,
\ee
where $\xi$, $\beta$ and $\lambda$ are dimensionless real positive
couplings, $h$ is the Higgs field in the unitary gauge (and we neglect
its present non-zero vacuum expectation value irrelevant for the
large-field dynamics), $g$ is
the determinant of the metric chosen as $ds^2=dt^2-a^2(t)d{\bx}^2$,  and the reduced Planck mass $M_P$
is defined via Planck mass $M_{Pl}$ as $M_P^2=M^2_{Pl}/(8\pi)$.  Upon
the Weyl transformation $g_{\mu\nu} \to g_{\mu\nu}\times \e^{\sqrt{\frac{2}{3}}\frac{\phi}{M_P}}$ 
with real scalar function $\phi(x)$ 
the action \eqref{action-1} takes form of the Einstein--Hilbert for
gravity and two coupled scalars -- Higgs $h$ and scalaron $\phi$ \cite{Ema:2017rqn,Gorbunov:2018llf}, 
\be
\label{action-4}
\begin{split}
  S=&\int d^4 x \sqrt{-g}\, \Biggl[ -\frac{M_P^2}{2}R +
  \frac{1}{2}\e^{-\sqrt{\frac{2}{3}}\frac{\phi}{M_P}}
  g^{\mu\nu}\d_{\mu}h\d_\nu h + \frac{1}{2}g^{\mu\nu}\d_{\mu}\phi\d_\nu \phi\\ &
  -\frac{1}{4}\e^{-2 \sqrt{\frac{2}{3}}\frac{\phi}{M_P}}\l\lambda
  h^4+\frac{M_P^4}{\beta}\l \e^{\sqrt{\frac{2}{3}}\frac{\phi}{M_P}}-1-\xi \frac{h^2}{M_P^2}\r^2\r\Biggr].
\end{split}
\ee
This form explicitly shows the absence of any strong coupling problems in
the scalar sector right up to the Planck scale if the model parameters
obey the inequality~\cite{Ema:2017rqn,Gorbunov:2018llf}
\be
  \label{perturbativity}
  \beta \gtrsim \frac{\xi^2}{4\pi}.
\ee
To describe the model dynamics during inflation another form of the action is more suitable. Changing the variables
$(h,\phi)\to (H,\Phi)$ according to
\[
  h \equiv \sqrt{6}M_P\,\e^{\frac{\Phi}{\sqrt{6}M_P}}\th\frac{H}{\sqrt{6}M_P},\qquad
 \e^{\frac{\phi}{\sqrt{6}M_P}}\equiv
 \frac{\e^{\frac{\Phi}{\sqrt{6}M_P}}}{\ch  \frac{H}{\sqrt{6}M_P}}, 
\]
one arrives at the following Lagrangian in the scalar sector
\begin{align}
  L & = \frac{1}{2}\ch^2{\l\frac{H}{\sqrt{6}M_P}\r} g^{\mu\nu}\d_\mu \Phi
    \d_\nu \Phi + \frac{1}{2} g^{\mu\nu}\d_\mu H \d_\nu H - V(\Phi,H),
  \label{action-kinetic} \\
  V & \equiv  
    \begin{array}[t]{l@{\,}l}
      9M_P^4\Bigg\{&\displaystyle
                    \lambda\sh^4{\l\frac{H}{\sqrt{6}M_P}\r} \\
                   &\displaystyle
                    +\frac{1}{36\beta}\left[ 1-e^{-\sqrt{\frac{2}{3}}\frac{\Phi}{M_P}}
                    \ch^2{\l\frac{H}{\sqrt{6}M_P}\r}-6\xi\sh^2{\l\frac{H}{\sqrt{6}M_P}\r}\right]^2
            \Bigg\}.
    \end{array}
  \label{action-potential}
\end{align}
The model exhibits effectively single-field inflation, whose trajectory in
the scalar field space is defined by nullifying the second term in the
potential \eqref{action-potential}, see
Refs.~\cite{Ema:2017rqn,Gorbunov:2018llf} for details. The matter
power spectrum normalization on the the Planck measurements
\cite{Ade:2015lrj} fixes the model parameters as \cite{Ema:2017rqn} 
\be
  \label{norm}
  \beta+\frac{\xi^2}{\lambda}\approx 2\times 10^{9}.  
\ee
The second term above must dominate to give the same predictions
for cosmological perturbations as in the original Higgs inflation,
which together with \eqref{perturbativity} constrain the viable range
of model parameters as
\be
\label{bounds}
\frac{\xi^2}{4\pi}<\beta < \frac{\xi^2}{\lambda}. 
\ee
The lower end of this range corresponds to ``Higgs-like'' behaviour (the scalar and tensor perturbation spectra are governed by parameters form the Higgs sector), while the upper end gets closer to $R^2$ inflation \cite{Starobinsky:1980te,Ema:2016dny,Gorbunov:2018llf}.
Recall that we treat $\lambda$ as a positive parameter taking its natural
value of $10^{-3}$--$10^{-2}$. Hence the tilt of scalar power spectrum
$n_s-1$ and the tensor-to-scalar ratio $r$ 
are as for pure-Higgs inflation \cite{Bezrukov:2007ep},  
\be
\label{R2-Higgs-predictions}
n_s=1-\frac{2}{N_e},\qquad r= \frac{12}{N_e^2},
\ee
with $N_e=50$--$60$ being a number of e-foldings until the end of
inflation. This number 
depends slightly on the reheating temperature, and our main task in
this paper is to get a reliable estimate of this quantity and hence to
refine the prediction \eqref{R2-Higgs-predictions} by obtaining an exact
value for $N_e$. 

At the onset of inflation $\Phi$ exceeds $H$ in value. The latter
remains almost constant, but then starts to evolve smoothly, and at $N_e\simeq
60$ before the end of inflation both fields crawl along the inflationary
valley, where the second term in scalar potential \eqref{action-potential} is zero. In the transverse direction the scalar potential is
always very steep, preventing the production of undesirable
isocurvature perturbations~\cite{Ema:2017rqn}. The equations of
motion for the homogeneous Higgs $H_0(t)$ and scalaron
$\Phi_0(t)$ fields follow from the action with Lagrangian
\eqref{action-kinetic}, \eqref{action-potential}:
\begin{align}
  \label{homogeneous-Phi}
  \ddot \Phi_0 + \l 3 {\cal H} +\sqrt{\frac{2}{3}} \frac{\dot H_0}{M_P} \tanh{\frac{H_0}{\sqrt{6}M_P}}\r \dot \Phi_0 +
  \frac{V_{\Phi_0}}{\cosh^2{\frac{H_0}{\sqrt{6}M_P}}}&=0\\
 \ddot H_0 + 3 {\cal H}\dot H_0 +
 V_{H_0}-\frac{\dot\Phi_0^2}{2\sqrt{6} M_P}\sinh{\frac{\sqrt{2}
     H_0}{\sqrt{3} M_P}}&=0.
 \label{homogeneous-H}
\end{align}
Here, $V_{H}$ and $V_{\Phi}$ stand for the potential derivatives with respect to corresponding fields, and $V_{\Phi_0,H_0}\equiv V_{\Phi,H}(\Phi_0,H_0)$. Dots denote time derivatives and the Hubble parameter ${\cal H}=\dot a/a$
is determined by the Friedman equation, as usual,
\begin{equation}
\label{Friedman}
  3 M_P^2 {\cal
    H}^2=\frac{1}{2}\dot\Phi_0^2\cosh^2{\frac{H_0}{\sqrt{6}M_P}} +
  \frac{1}{2}\dot H_0^2 + V(H_0,\Phi_0).
\end{equation}
By the end of
inflation both scalar homogeneous fields are sub-Planckian~\cite{Gorbunov:2018llf}, 
\begin{equation}
  \label{limits-on-fields-from-above}
|\Phi_0|\lesssim 0.3M_P, \qquad |H_0|\lesssim 0.03M_P,
\end{equation}
and actively participate in the model dynamics. The Hubble scale at the end of inflation can be estimated from the analysis of Ref.~\cite{Ema:2016dny} as 
\begin{equation}
    \label{Hubble-end}
    {\cal H} = \frac{\sqrt{\lambda}}{3\sqrt{2}\xi}\Phi_0< 0.1\frac{\sqrt{\lambda}}{\xi} M_P.
\end{equation}
First, one can check that the smallness of $H_0$ \eqref{limits-on-fields-from-above} guarantees that terms following from the non-canonical kinetic term in \eqref{action-kinetic} 
can be neglected in equation of motions \eqref{homogeneous-Phi} \eqref{homogeneous-H}.  Then, 
in the small-field regime, the scalar potential \eqref{action-potential} can be
approximated as polynomial of the fourth order in fields,
\begin{equation}
  \label{quartic-potential}
\begin{split}
  V(H, \Phi) & =\frac{1}{4}\l \lambda +\frac{\xi^2}{\beta}\r H^4
  + \frac{M_P^2}{6\beta}\Phi^2 - \frac{\xi M_P}{\sqrt{6}\beta}\Phi H^2\, \\
&+ \frac{7}{108\beta}\Phi^4 + \frac{\xi}{6\beta}\Phi^2 H^2 -
  \frac{M_P}{3\sqrt{6}\beta}\Phi^3.
\end{split}
\end{equation}
Hereafter we take the leading order terms in $\xi\gg 1$.
Right after inflation the
potential \eqref{quartic-potential} refers to a system of two
coupled and highly anharmonic oscillators, whose initial effective
frequencies are not far from the Planck scale, but rapidly decrease as
the Universe expands and the field amplitudes fall down. The scalar potential is symmetric with respect to a change of the sign of the Higgs field, but contains terms odd in $\Phi$. The last term in the first line of \eqref{quartic-potential} is of the special interest, as it can give a large negative contribution (dominating with $\xi\gg 1$ over the negative contribution of the last term in second line of \eqref{quartic-potential}). 

Note in passing that the first two terms in the first line turn into the SM Higgs self-coupling and scalaron mass terms, which dominate the potential at very small fields. One recognises that for our range of parameters \eqref{bounds}, not $\lambda$, but mostly $\xi^2/\beta$ defines the Higgs self-coupling constant at large sub-Planckian fields. However, below the energy scale of the scalaron mass $\Phi$ must be integrated out at the quantum level. As a result, the last term in the first line of \eqref{quartic-potential} provides a contribution to the Higgs self-coupling which exactly cancels the $\xi^2/\beta$-part below this scale, making the parameter $\lambda$ solely responsible for the Higgs self-coupling at low energies, as it must be in the SM. In principle, to verify the model, one should independently measure all the parameters in particle collisions -- the Higgs self-coupling (or mass, which has been already done), scalaron mass $M_P/\sqrt{3\beta}$, and scalaron-Higgs interaction (which would allow to measure $\xi$).  Then, independently, CMB observations give the relation \eqref{norm}, which could then be tested against the measurements in particle collisions.

Let us characterise the motion of the uniform background fields $\Phi_0$ and $H_0$ after inflation.
Typically one has $|\Phi_0|>|H_0|$, and the timescale corresponding to the oscillations in $\Phi_0$ is much longer than for $H_0$ direction.  Therefore one can split the dynamics into periods with positive and negative $\Phi_0$. 

For $\Phi_0<0$ all terms in the potential \eqref{quartic-potential} are positive, and $\Phi_0$ makes half-oscillation with a typical frequency not lower than the scalaron mass, 
\[
  \left.\omega^2_{\Phi_0}\right|_{\Phi_0<0} > M_P^2/3\beta,
\]
which follows from the second term of \eqref{quartic-potential}.
At the same time the field $H_0$ oscillates about the origin with frequency depending on $\Phi_0$. The oscillation type (harmonic or anharmonic) depends on the Higgs amplitude and $\Phi_0$ -- i.e.\ which of the terms in \eqref{quartic-potential} dominates. In the harmonic regime, one can read the frequency from \eqref{quartic-potential} as 
\begin{equation}
  \label{H-f<0}
  \left.\omega_{H_0} ^2 \right|_{\Phi_0 <0} = - \frac{\sqrt{2}\xi}{\sqrt{3}\beta}M_P\Phi_0.
\end{equation} 
Note, that since the Higgs evolution timescale is much shorter than that of the scalaron, the Higgs contribution can be averaged over its oscillations, further increasing the scalaron effective frequency.

The case of positive $\Phi_0$ is the most interesting for particle production.  Here the fields evolve along the valley where the second term in potential \eqref{action-potential} is zero ($\Phi_0 ^2$ and $\Phi_0 H_0 ^2$ terms in \eqref{quartic-potential} cancel each other),
\begin{equation} \label{hmin}
  H_{\mathrm{min}}^2 \equiv \frac{2\xi}{\sqrt{6}(\xi^2+\lambda\beta)} M_P\Phi_0
                     \simeq \sqrt{\frac{2}{3}}\frac{M_P\Phi_0}{\xi},
\end{equation}
where in the last equality we used the assumption \eqref{bounds}. 
The fields oscillate around this trajectory, which departs from the origin, evolves towards large fields, then stops and returns back to the next zero crossing of the scalaron. The typical scalaron timescale of the evolution along the valley and back can be estimated by replacing the $H_0$ in the quartic potential \eqref{quartic-potential} by its value at the bottom of the valley $H_{\mathrm{min}}$ and calculating the effective quadratic part in $\Phi_0$. This gives for the scalaron frequency squared in the harmonic regime
\begin{equation}
  \left.\omega_{\Phi_0} ^2 \right|_{\Phi_0 >0}
  = \frac{M_P ^2}{3}\frac{\lambda}{\xi^2 + \lambda \beta}
  \simeq \frac{\lambda M_P ^2}{3\xi^2},
\end{equation}
which is lower, see \eqref{bounds}, than that at $\Phi_0<0$, limited from below by the scalaron frequency in  $R^2$-inflation~\cite{Starobinsky:1980te}, $M_P^2/(3\beta)$, and actually corresponds better to the frequency of oscillations in pure Higgs inflation, see Refs.\cite{Bezrukov:2008ut,GarciaBellido:2008ab}. 
At large $\Phi_0$ the higher order terms in \eqref{quartic-potential} can make the oscillations slightly anharmonic. Finally, the frequency of oscillations of $H$, transverse to the valley, can be found from the second derivative of the potential \eqref{quartic-potential} $V_{HH}$ calculated in a point along the valley floor \eqref{hmin} 
\begin{equation}\label{omegaH}
  \left.\omega_{H_0}^2\right|_{\Phi_0 >0} = 2\sqrt{\frac{2}{3}}\frac{\xi M_P}{ \beta }\Phi_0.
\end{equation}
Note, this frequency parametrically coincides with that at negative $\Phi_0$ \eqref{H-f<0}. 

One observes, that at $\Phi_0>0$ there are two equivalent valleys in the potential, for positive and negative \emph{Higgs} values, $H_0>0$ and $H_0<0$.  When the scalaron $\Phi_0$ returns from negative steep potential to the positive valleys, the system can end up in either of them.  The choice depends on the number of oscillations that $H_0$ made while the scalaron field was completing its half oscillation at $\Phi_0<0$.  This ratio is determined by the model parameters only 
(values of $\lambda$, $\xi$, and $\beta$), but in practice is 
impossible to be estimated analytically, as far as it depends on the nonlinearities of the potential \emph{and} on the amplitudes of the fields in the given oscillation.  Indeed, while for the first scalaron zero crossing after the end of inflation the ``choice'' of the valley can be found analytically given the smooth ending of inflation, for further oscillations the situation is more complicated, as far as the system approaches zero not simply along the valley \eqref{hmin}, but with some oscillations in the $H_0$ direction.

\begin{figure*}[htb]
    \centering
    \includegraphics[width=0.33\linewidth]{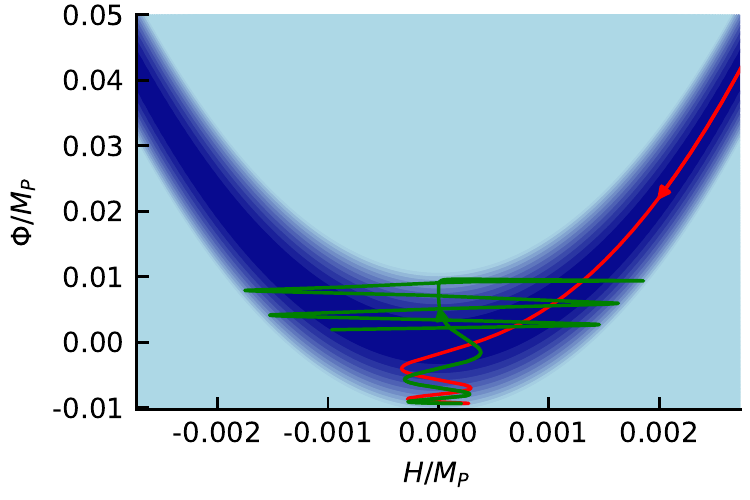}%
    \includegraphics[width=0.33\linewidth]{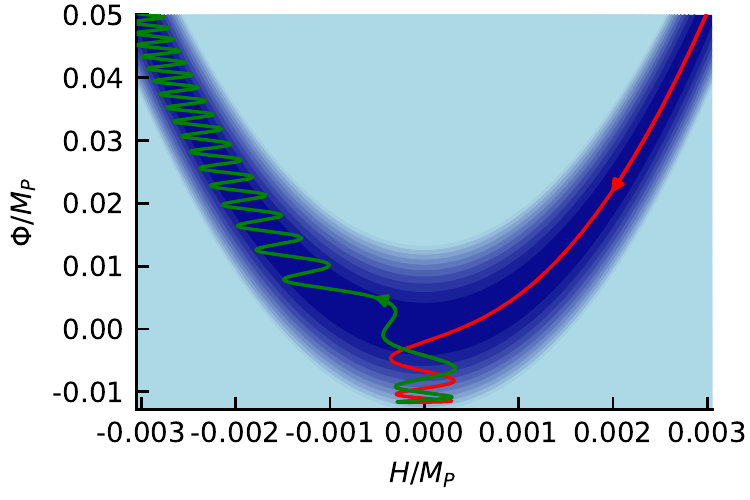}%
    \includegraphics[width=0.33\linewidth]{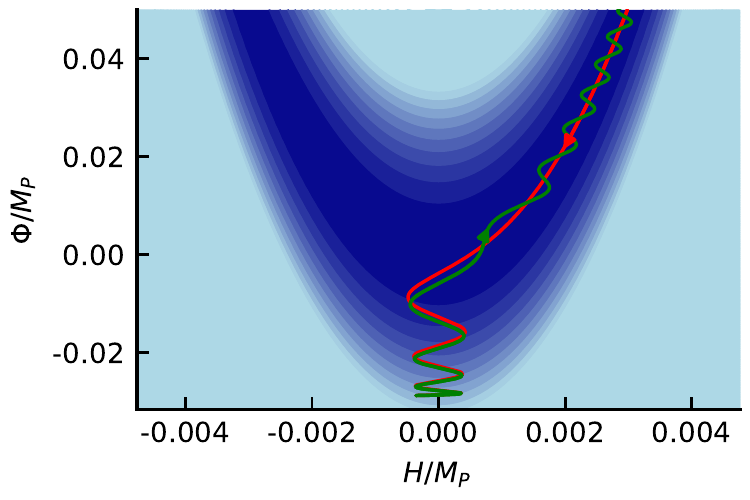}\\
    \includegraphics[width=0.33\linewidth]{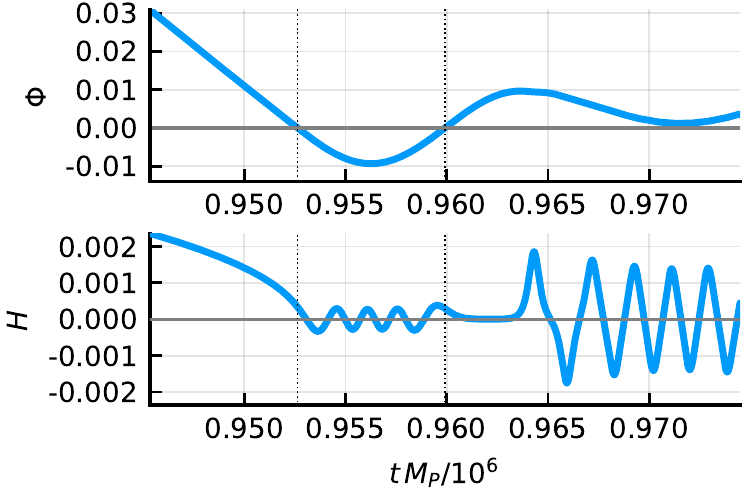}%
    \includegraphics[width=0.33\linewidth]{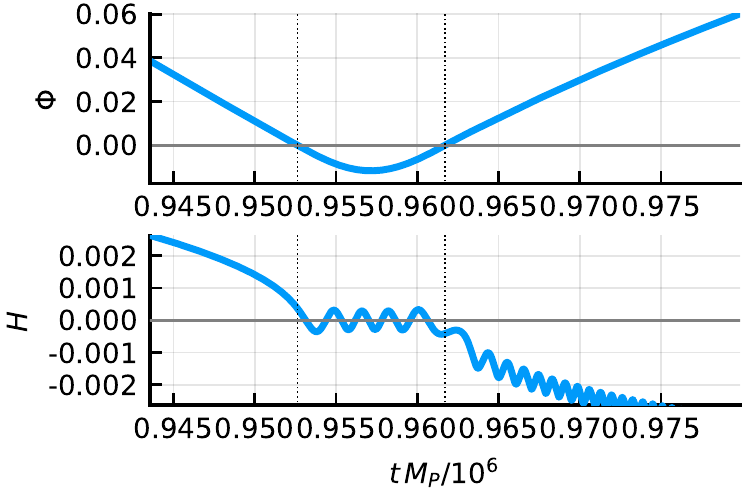}%
    \includegraphics[width=0.33\linewidth]{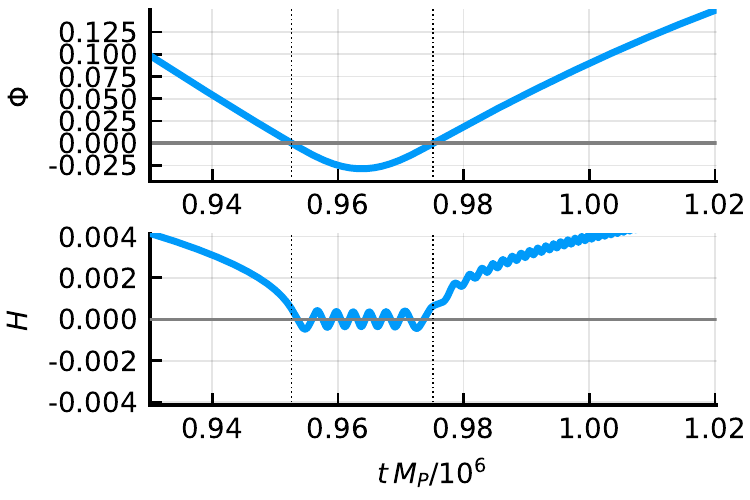}
    \caption{Illustration of the field evolution after inflation around first zero crossing.  The shading on the plot reflects the shape of the  potential.  Red and green colours for the field trajectory correspond to negative and positive $\dot{\Phi}_0$. Different types of behaviour can be seen for $\beta=(1.869,2.9,18.0)\times10^6$ from left to right panel. Thin dashed vertical lines mark the $\Phi_0=0$ moments.}
    \label{bgZoom}
\end{figure*}

There is a bifurcation point, where the background Higgs field is displaced precisely the right amount at the time of scalaron zero crossing to neither overshoot nor undershoot the local potential maximum $H_0 = 0$ for positive scalaron values (see Fig.~\ref{bgZoom}, left plot).   This situation is rather peculiar -- the exact borderline between the two stable trajectories corresponds to unstable motion with $H_0=0$ and growing $\Phi_0>0$ -- or the fields ``climbing'' up along the middle of the barrier separating the $H_0 <0$ and $H_0 >0$ valleys.  It can be immediately seen that for such a configuration the massive term for the Higgs field in \eqref{quartic-potential} (third term) is tachyonic.  
This is exactly the source of the tachyonic instability that we study in detail in the next section.  The closer the trajectory of the background fields is to the bifurcation situation, the longer the system stays in the tachyonic region, leading to more efficient decay of the background into inhomogeneous modes.  Also, bifurcation points correspond to the maximum transfer of the energy from the motion in the $\Phi_0$ direction into the oscillations in the $H_0$ direction.

We conclude the study of homogeneous field dynamics with  the
observation that in this oscillatory picture, the Friedman equation shows that for (\ref{limits-on-fields-from-above}) the Hubble timescale is much larger than a period of $\Phi_0$ oscillation, which in turn is much larger than a period of Higgs oscillations. Over several $\Phi_0$ periods, the expansion of the Universe can safely be ignored.

\section{Equation of motions for the inhomogeneous modes}
\label{Sec:III}

As the two coupled scalars \eqref{quartic-potential} begin to oscillate, they produce particles and eventually can reheat the Universe. Since particle production in pure $R^2$-inflation \cite{Starobinsky:1980te,Vilenkin:1985md,Gorbunov:2010bn} exploits very slow gravitational dynamics, rather than the fast electroweak dynamics that cause reheating in pure Higgs inflation \cite{Bezrukov:2008ut,GarciaBellido:2008ab,Sfakianakis:2018lzf}, one expects the Higgs homogeneous fields to be responsible for the early-time reheating in the mixed Higgs-$R^2$ case. 

Particle production in the scalar sector (scalarons and Higgs bosons)
is described by the following quadratic action (derived
from \eqref{action-kinetic} and \eqref{action-potential}) for
inhomogeneous perturbations $\tilde H(t,{\bx})$ and $\tilde \Phi
(t,{\bx})$,
\begin{equation}
  \label{quadratic-action}
  \begin{split}
    S^{(2)}\!=\!\int\!\! \sqrt{-g}\,d^4x \Biggl[
    \half g^{\mu\nu}\d_\mu\tilde H \d_\nu \tilde H + \half
    \cosh^2{\frac{H_0}{\sqrt{6} M_P}} g^{\mu\nu}\d_\mu \tilde\Phi \d_\nu
    \tilde\Phi + \frac{\dot \Phi_0}{\sqrt{6}M_P}
    \sinh{\frac{2H_0}{\sqrt{6}M_P}} \dot{\tilde\Phi}\tilde H \\
  +\frac{\dot\Phi_0^2}{12M_P^2}\cosh{\frac{2H_0}{\sqrt{6}M_P}} \tilde H^2
    -\half V_{\Phi_0\Phi_0}\tilde\Phi^2 - V_{\Phi_0 H_0} \tilde \Phi
    \tilde H - \half V_{H_0 H_0}\tilde H^2
  \Biggr].
  \end{split}
\end{equation}
As discussed in Sec.\,\ref{Sec:II}, at small homogeneous fields we set $\cosh\left({H_0}/{\sqrt{6} M_P}\right) \simeq 1$, reducing the first four terms in \eqref{quadratic-action} to a pair of canonical kinetic terms. The second derivatives of the scalar potential are well approximated by (see eq.\,\eqref{quartic-potential}) 
\begin{align}
  \label{mass-1}
  V_{\Phi_0\Phi_0}&=\frac{1}{3\beta}M_P^2+\frac{\xi}{3\beta}H_0^2-\frac{\sqrt{2}}{\sqrt{3}\beta}
  M_P\Phi_0 + \frac{7}{9\beta}\Phi_0^2\\
  \label{mass-2}
  V_{\Phi_0H_0}&=-\frac{\sqrt{2}\xi}{\sqrt{3}\beta} M_P H_0
  +\frac{2\xi}{3\beta}\Phi_0 H_0\\
  \label{mass-3}
  V_{H_0H_0}&=3\l \lambda+\frac{\xi^2}{\beta}\r H_0^2 -
  \frac{\sqrt{2}\xi}{\sqrt{3}\beta} M_P\Phi_0 + \frac{\xi}{3\beta}\Phi_0^2
\end{align}
In the above expressions all terms linear in fields become negative when $\Phi_0$ gets positive value and can potentially lead to tachyonic
instabilities -- as we discuss in the next section.

To the leading order in background fields one can obtain from the
quadratic action \eqref{quadratic-action} the equations of motion for
the Fourier transforms, $H_k$ and $\Phi_k$
of the Higgs and scalaron inhomogeneous modes,
\begin{align} \label{elOne}
  \ddot H_k + 3 {\cal H}\dot H_k + \frac{k^2}{a^2}H_k +  V_{H_0 H_0}  H_k +V_{\Phi_0 H_0} \Phi_k  &=0\\
  \ddot \Phi_k + 3 {\cal H} \dot \Phi_k  + \frac{k^2}{a^2} \Phi_k + V_{\Phi_0
    \Phi_0}\Phi_k\ +V_{\Phi_0 H_0} H_k
 \label{elTwo}
 &=0.
\end{align}
Using \eqref{mass-1}-\eqref{mass-3}, these equations to leading order are
\begin{align} \label{elOneS}
  \ddot H_k + \l\frac{k^2}{a^2} +3\l \lambda+\frac{\xi^2}{\beta}\r H_0^2-\sqrt{\frac{2}{3}}\frac{\xi}{\beta}M_P\Phi_0\r H_k-\sqrt{\frac{2}{3}}\frac{\xi}{\beta}M_PH_0 \Phi_k &=0
  \\
  \ddot \Phi_k + \l \frac{k^2}{a^2}  + \frac{1}{3\beta}M_P^2 +\frac{\xi}{3\beta}H_0^2\r \Phi_k -\sqrt{\frac{2}{3}}\frac{\xi}{\beta}M_PH_0 H_k
  &=0. \label{elTwoS}
\end{align}
They describe two linearly coupled oscillators with time-dependent mass terms, which may lead to particle production. When $\Phi_0<0$ both diagonal terms are positive, and the off-diagonal terms are not large enough to make the squared frequency negative, initiating an instability in the system. However, the mass terms rapidly oscillate with the background Higgs field, potentially producing particles. At the positive branch, $\Phi_0>0$, the heavy (Higgs-like) mass eigenstate can become tachyonic. Indeed, in the potential valley with the homogeneous Higgs mode in the minimum \eqref{H-f<0}, the term in the parenthesis in \eqref{elOneS} reduces to 
\begin{equation}
\label{wH-in-minima}
\omega_H^2({\bf k})=\frac{k^2}{a^2}+2\l \lambda+\frac{\xi^2}{\beta}\r H_{\mathrm{min}}^2.
\end{equation}
Higgs oscillations about $H_{\mathrm{min}}$ with sufficiently large amplitudes give a contribution to \eqref{wH-in-minima} which may turn it negative. This induces the tachyonic instability meaning instant particle production. Note that production is more efficient at small $H_{\mathrm{min}}$, where the system just left the bifurcation point and started to evolve along the potential valley. Later $H_{\mathrm{min}}$ becomes large enough to prevent this term flipping sign because of the Higgs oscillations. Anyway, the larger the oscillation amplitude, the more efficient the production. The source is the homogeneous Higgs field, so the outcome is constrained not only by the Higgs amplitude (which defines the highest momentum of produced particles saturating the total density and energy of produced particles) but most severely by the amount of total energy concealed in the Higgs field. 

For the weak gauge bosons the mass terms are determined by the Higgs
fields \cite{Gorbunov:2018llf}. The quadratic Lagrangian for the $W^\pm$-bosons reads
\[
L_g^{(2)}= - \frac{1}{2}\l \d_\mu W^+_\nu - \d_\nu W^+_\mu\r  
\l \d_\lambda W^-_\rho - \d_\rho W^-_\lambda\r g^{\mu\lambda}g^{\nu\rho}  
+\frac{g^2 H_0 ^2}{4}\, W_\mu^+ W_\nu^-g^{\mu\nu},
\]
where $g$ is the weak gauge coupling constant; in the very small field limit, $H_0\to v=246$\,GeV, we restore here the SM mass term of $W^\pm$-bosons. In what follows we consider the
$W^\pm$-bosons only, the case of $Z$-bosons is similar.  Defining the
field-dependent variable
\begin{equation}
  \label{transverse-mass}
m_T \equiv  \frac{g}{2} H_0,
\end{equation}
one obtains for the Fourier modes of transverse components of
$W$-bosons a damped Klein-Gordon equation
with time-dependent mass:
\begin{equation}\label{elW}
\ddot W_k^T + 3{\cal H} \dot W_k^T + \frac{k^2}{a^2}W_k^T + m_T^2 W_k^T=0.
\end{equation}
While the last term in the equation above, being rapidly oscillating,  generically sources the
particle production, in our case with functional form
\eqref{transverse-mass} and rather small Higgs field immediately after
inflation the production is not very efficient, see Ref.~\cite{He:2018mgb} 
for details.

However, the situation is different for the longitudinal modes, which 
also obey the Klein-Gordon equation
\begin{equation}
  \ddot{W}_k ^L + 3\mathcal{H} \dot{W}_k ^L + \omega_W ^2 (\mathbf{k})W_k ^L = 0.
  \label{EqWl}
\end{equation} 
with frequency (see Ref.~\cite{Ema:2016dny} for its conformal analog)  
\begin{equation} \label{omegawexact}
\omega^2_W (\mathbf{k})=\frac{k^2}{a^2}+m_T^2- \frac{k^2}{k^2+a^2m_T^2}
\l \dot {\cal H}+2{\cal H}^2+3{\cal H}\frac{\dot m_T}{m_T}+\frac{\ddot
m_T}{m_T}- \frac{3(\dot m_T+{\cal H}m_T)^2}{k^2/a^2+m_T^2}\r.
\end{equation}
In case of large 3-momenta $k/a\gg m_T$, expression \eqref{omegawexact} reduces to
\begin{equation} \label{omegaW}
  \omega_W^2 \approx \frac{k^2}{a^2} + g^2 \frac{
      H_0 ^2}{4}+ \frac{V_{H_0}}{H_0} - \frac{2}{3 M_P ^2 } V (H_0,\Phi_0)
   \equiv \frac{k^2}{a^2}+m_W^2(k/a\gg m).
\end{equation} 
For the fields obeying \eqref{limits-on-fields-from-above} one obtains the leading contributions 
\begin{equation}\label{wwSimple}
  \omega_W^2 = \frac{k^2}{a^2}+\frac{g^2}{4}H_0^2 + \frac{\xi}{3\beta}\Phi_0^2 + \l \lambda +\frac{\xi^2}{\beta}\r H_0^2 - \frac{\xi \sqrt{2}}{\beta\sqrt{3}} M_P\Phi_0\,,
\end{equation}
where the last two terms cancel along the valley, and the Higgs oscillations about its potential minimum with sufficiently large amplitude of $H_0 - H_{\mathrm{min}}$, see \eqref{hmin},  close to a scalaron zero crossing can give a dominant tachyonic contribution.

This post-inflationary feature is responsible for the
violent production of longitudinal modes of the weak gauge bosons and 
instant preheating in the model, as we show in the next sections.


\section{Description of particle production}
\label{Sec:IV}

In order to study the enhancement of the inhomogeneous scalar modes, they must be quantized in a canonical manner. To do so, we first remove the determinant of the metric in (\ref{quadratic-action}) by rescaling fields as
\(
  \tilde{\Phi}(t,\mathbf{x}) \equiv a^{3/2}\Phi(t,\mathbf{x}),\) \(
  \tilde{H}(t,\mathbf{t})\equiv  a^{3/2}H(t,\mathbf{x}),
\)
what removes the Hubble friction term in \eqref{elOne}, \eqref{elTwo} and adds irrelevant Hubble suppressed term to the mass.
To diagonalise the mass matrix we rotate $(\tilde{\Phi},\tilde{H})\to(\varphi_L,\varphi_H)$ with time-dependent mixing angle $\theta$
\begin{equation}\label{tTheta}
  \tan 2 \theta = \frac{2 V_{\Phi_0 H_0}}{V_{\Phi_0 \Phi_0} - V_{H_0 H_0}}.
\end{equation}
The resulting mass eigenvalues are
\begin{equation}\label{eigenvalues}
m_{L,H}^2 = \frac{1}{2} \left( V_{H_0 H_0} + V_{\Phi_0 \Phi_0} \right) \\ \times 
\left( 1 \pm \sqrt{1 - 4 \frac{V_{\Phi_0 \Phi_0} V_{H_0 H_0}  - V_{\Phi_0 H_0} ^2}{\left( V_{H_0 H_0}  + V_{\Phi_0 \Phi_0} \right)^2}} \right) .
\end{equation}
Except for short moments around $\Phi_0\approx0$ one is much heavier than the other, corresponding to the oscillations in transverse directions of the potential valley and along the valley. To qualitatively understand the reheating mechanism we can focus just on the heavy inhomogeneous Higgs-like mode. To leading order, its mass equals, see eq.~\eqref{mass-3},  
\begin{equation}\label{mHs}
m_{H,L}^2\approx V_{H_0H_0}\approx 3\l \lambda+\frac{\xi^2}{\beta}\r H_0^2-\frac{\sqrt{2}\xi}{\sqrt{3}\beta}M_P\Phi_0.
\end{equation}
This expression gives the estimate for the mass mode with the \emph{largest} absolute value, and coincides with $m_H^2$ for positive mass, and $m_L^2$ for tachyonic mass.
It coincides with the diagonal mass of the mode $H_k$ in \eqref{elOneS}. At $\Phi_0<0$ it is positive, while can be negative for positive scalaron. The mode frequency at the minimum along the potential valley equals \eqref{wH-in-minima}, which allows for negative values with sufficiently large amplitude of the Higgs oscillations, as explained right after eq.~\eqref{wH-in-minima}. We confirm this numerically in Sec.~\ref{Sec:V}.

The diagonal modes have physical frequencies 
\begin{equation}
\omega_{L,H}(\mathbf{k})^2 \equiv \frac{k^2}{a^2} + m_{L,H}^2.
\end{equation}
In the Hamiltonian there are terms proportional to $\dot{\theta}$, which rotate fields and momenta into one another over a timescale $\sim \dot{\theta}^{-1}$. This decoherence process has been studied extensively in \cite{Polarski:1995jg}.  
Numerically one finds that $\dot{\theta}^2 \ll \left| m_L ^2 \right|$ at all times, except during a short region of time near $\Phi_0 = 0$, where the mode roles exchange. Outside of this region, the $L$ and $H$ modes are separable. Furthermore, when $\omega_L$ and $\omega_H$ are varying adiabatically slowly, one has WKB solutions to (\ref{elOneS}) and (\ref{elTwoS}), and a particle interpretation for $\varphi_{(L,H)}$. This is true for any time when the scalaron is sufficiently far away from $\Phi_0 = 0$, so one may legally quantize $\varphi_{(L,H)}$ during these adiabatic time slots. 
The particle production can be obtained from the solutions to eqs.~\eqref{elOne}, \eqref{elTwo} (and, in a similar way for \eqref{EqWl}) subject to the vacuum initial conditions
$ f_\bk(t)=\e^{-i\omega t}/\sqrt{2\omega(\bk)}$ at $t\to 0$, 
where $f$ stands for each of the $\varphi_H$, $\varphi_L$, $W^L$, $W^T$ with the respective frequency.
Taking these solutions for each mode at large $t$ when the system comes back to adiabatic behaviour, one obtains
the comoving number density per phase space volume ${d^3\bk}/{(2\pi)^3}$ of  particles of each type \cite{Polarski:1995jg}
\begin{equation} \label{nDef}
  n_\bk = \frac{1}{2} \left| \sqrt{\omega (\mathbf{k})} f_\bk - \frac{i}{\sqrt{\omega (\mathbf{k})}} \dot{f_\bk}\right|^2
\end{equation}
and the physical energy density of each species is
\begin{equation} \label{rhoPhys}
  \rho = \int \frac{d^3\bk}{(2\pi)^3a^3(t)} \omega(\mathbf{k}) n_\bk.
\end{equation}

\section{Numerical study and results}
\label{Sec:V}

To study the preheating process numerically we set $\lambda = 0.01$ and consider three particularly illustrative values of $\beta$: $\beta\equiv(\beta_1,\beta_2,\beta_3)=(1.869,2.9,18.0)\times10^6$. The first value corresponds exactly to the system crossing the first zero in the previously-discussed bifurcation regime, where the background Higgs oscillates right about zero for an extended period, while the scalaron oscillations are nearly trapped in the vicinity of zero. Value $\beta_2$ is somewhat off the tuned value, the background fields rapidly return to one of the positive $\Phi_0$ potential valleys with significant oscillations on present in the $H_0$ direction. Finally, $\beta_3$ is far away from any special points, the Higgs oscillation amplitude along the valley floor is small.\footnote{There are also special values of $\beta$ where the system reflects into the valley exactly along $H_{\mathrm{min}}(\Phi_0)$.} The evolution of the background fields near the first scalaron zero crossings is shown for these cases in Fig.~\ref{bgZoom}.

\begin{figure*}[htb]
	\centering
	\includegraphics[width=0.333\textwidth]{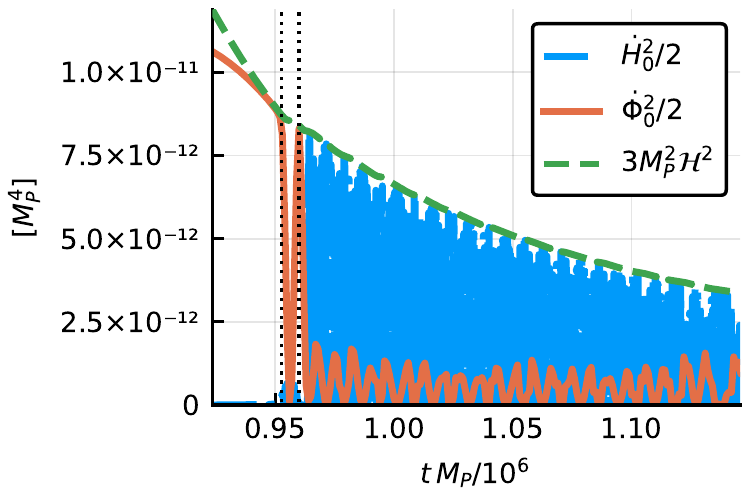}%
	\includegraphics[width=0.333\textwidth]{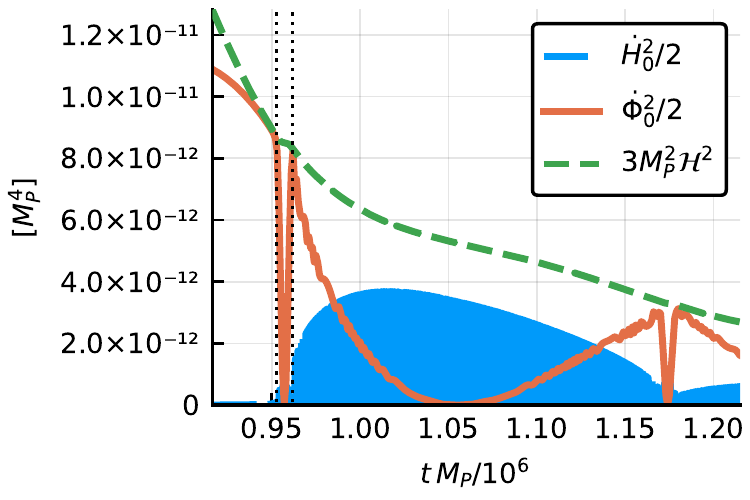}%
	\includegraphics[width=0.333\textwidth]{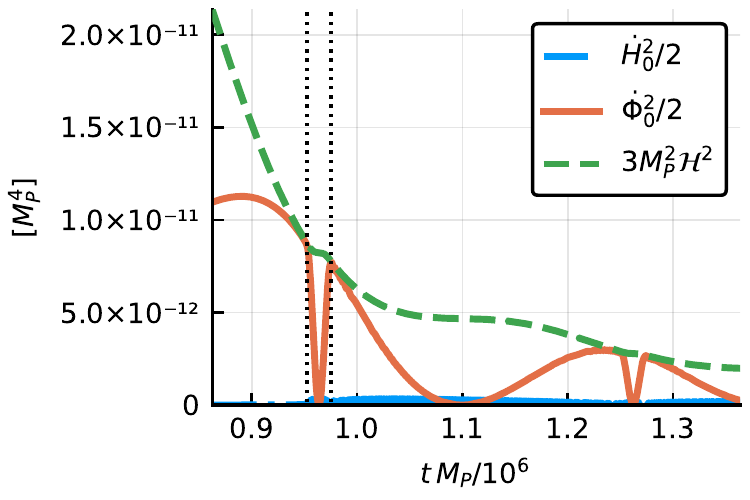}
	\caption{$ \dot{H}_0 ^2$, $\dot\Phi_0^2$ and the total background energy density for $\beta_1$, $\beta_2$ and $\beta_3$ from left to right.}
	\label{bgPlots}
\end{figure*}

\begin{figure*}[htb]
	\centering
    \includegraphics[width=0.333\linewidth]{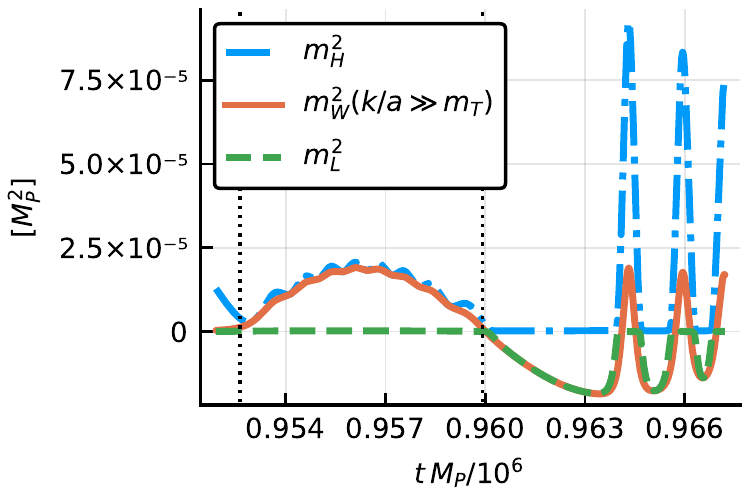}%
    \includegraphics[width=0.333\linewidth]{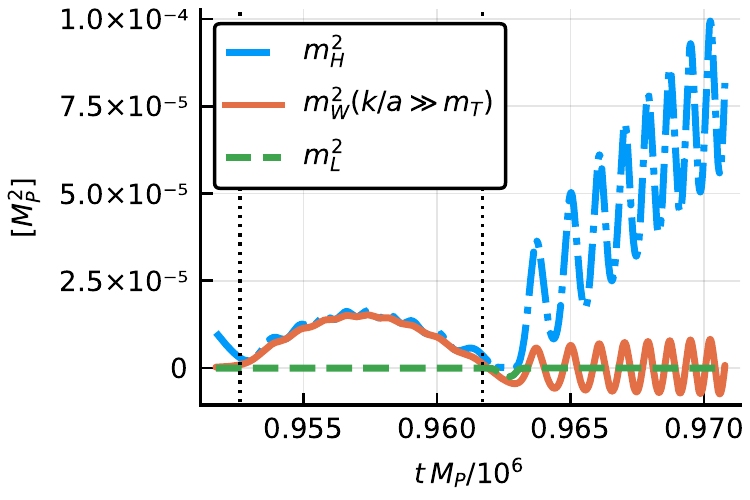}%
    \includegraphics[width=0.333\linewidth]{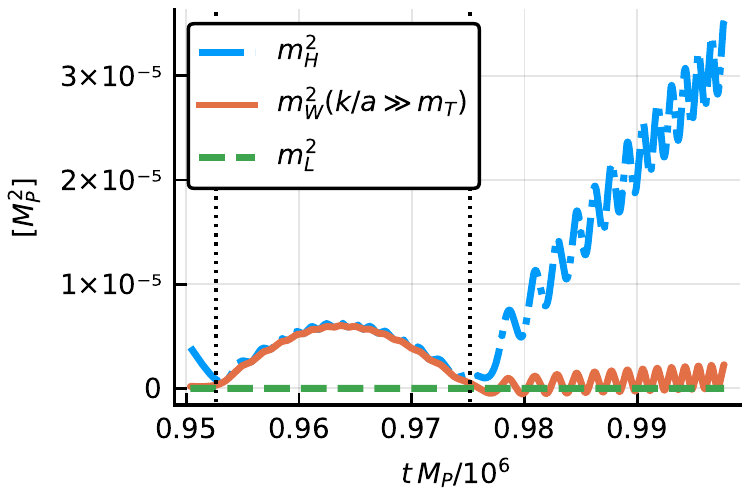}%
	\caption{Mode masses around the first zero-crossing of $\Phi_0$, for $\beta_1$, $\beta_2$ and $\beta_3$ from left to right.}
	\label{mPlots}
\end{figure*}

Consequently, the homogeneous Higgs gets the greatest fraction of the total energy in the first scenario, and the least in the third, see Fig.~\ref{bgPlots}.
The corresponding masses of the inhomogeneous Higgs and longitudinal components of the weak bosons are plotted around the first zero-crossing of $\Phi_0$ in Fig.~\ref{mPlots}.
As predicted by \eqref{wwSimple} and \eqref{mHs}, when the background Higgs mode is less energetic, the masses of the inhomogeneous modes are not allowed to oscillate as deeply into the tachyonic regime.

To study the preheating dynamics, the mode functions for $W_L$, and $(L,H)$ scalar perturbations were evolved numerically according to (\ref{elOne}), (\ref{elTwo}) and (\ref{EqWl}), subject to the vacuum initial conditions. The total energy in the inhomogeneous excitations of each type was calculated using \eqref{rhoPhys}.
This calculation is valid only while the backreaction on the uniform background can be neglected (or, roughly, while the energy density of the produced perturbations is below the uniform background energy density).

\begin{figure*}[htb]
	\centering
    \includegraphics[width=0.33\linewidth]{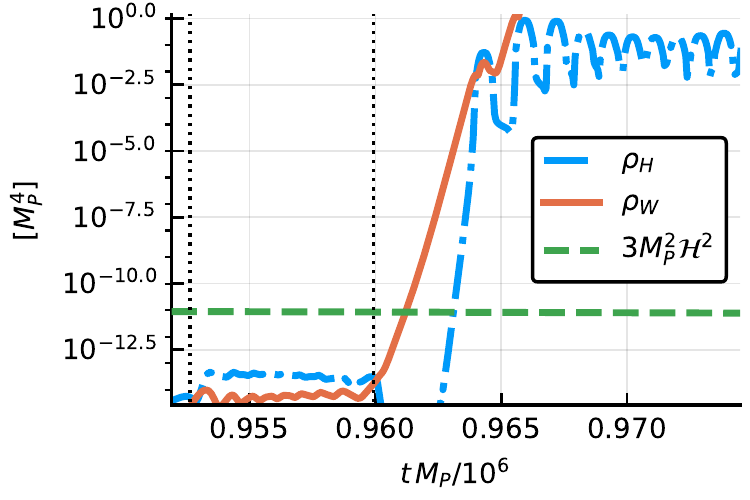}%
    \includegraphics[width=0.33\linewidth]{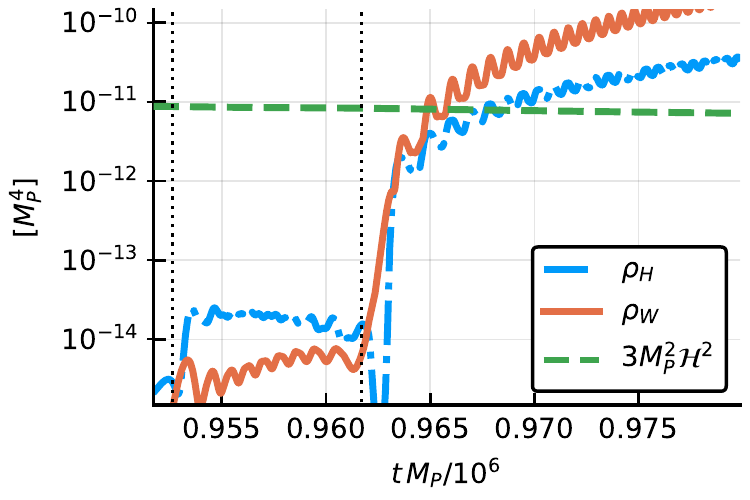}%
    \includegraphics[width=0.33\linewidth]{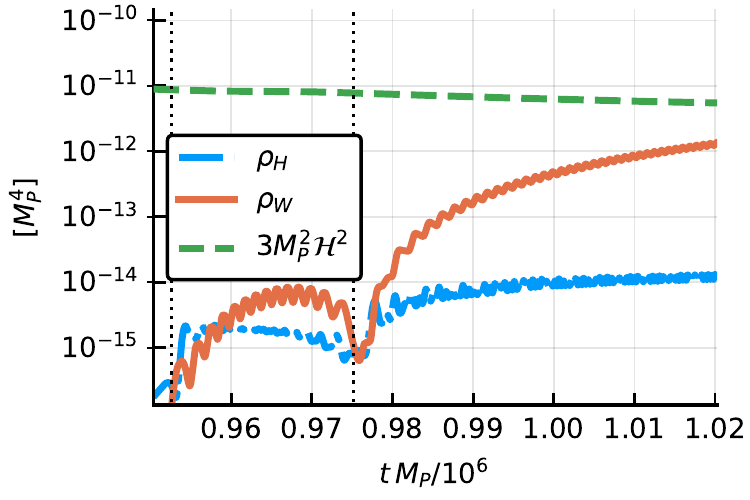}%
	\caption{
		Evolution of energy densities for $\beta_1$, $\beta_2$ and $\beta_3$  from left to right.
	}
	\label{rhoPlots}
\end{figure*}

In Fig.~\ref{rhoPlots} the physical energy density is plotted for both the Higgs perturbations and $W_L$, and compared to the background energy density $3 {\cH}^2M_P^2$. 
In the first two plots, the background Higgs energy is drained more or less instantaneously. However, parameters further away from the tachyonic situation lead to a significantly slower rise in the corresponding energy densities (third plot).

\begin{figure*}[htb]
	\centering
	\includegraphics[width=0.33\textwidth]{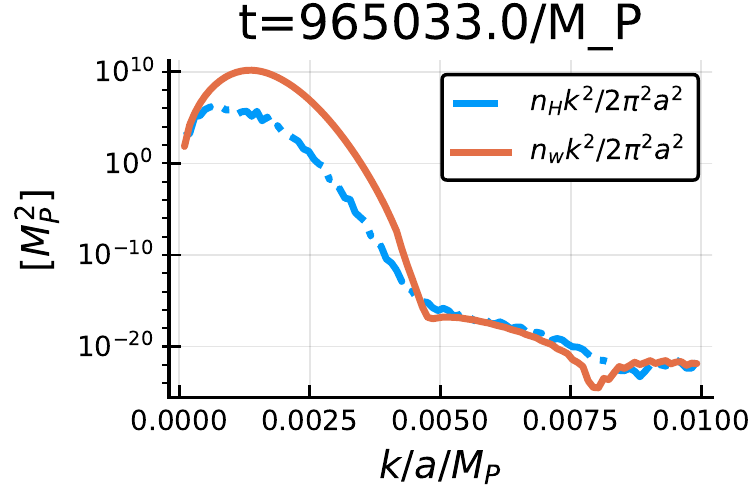}%
	\includegraphics[width=0.33\textwidth]{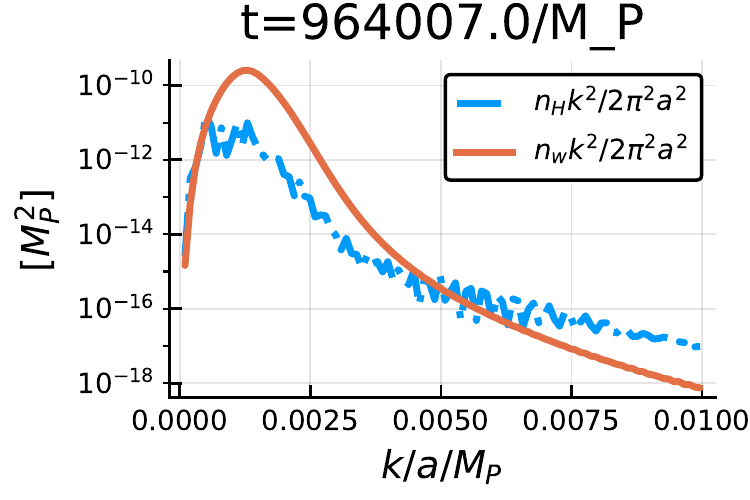}%
	\includegraphics[width=0.33\textwidth]{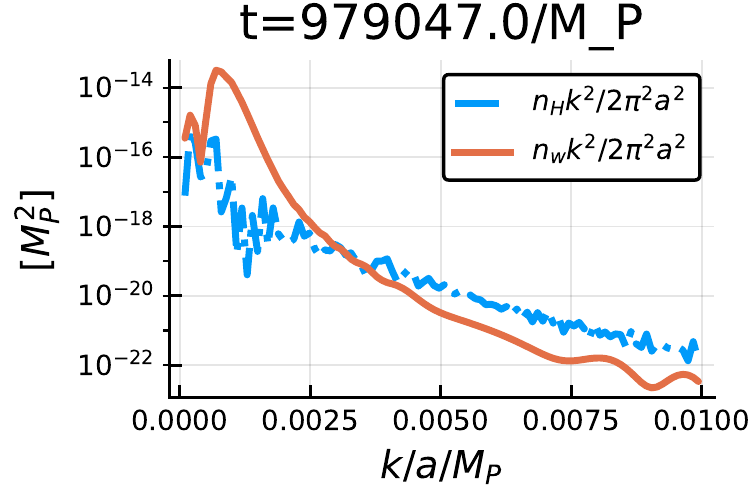}%
	\caption{Spectra of produced modes immediately after the tachyonic region for $\beta_1$, $\beta_2$ and $\beta_3$.}
	\label{spectraPlots}
\end{figure*}
It is insightful to investigate the spectrum for each situation. In Fig.~\ref{spectraPlots} the number density of each mode is plotted immediately after the tachyonic region for $\beta_1$ and $\beta_2$, and at a comparable point in the scalaron oscillation for $\beta_3$.  The typically produced particles are sharply peaked at low momenta, where the frequencies become tachyonic after the scalaron zero crossing. All species are non-relativistic.  A warning is in order -- as far as the analysis of the current paper does not account for the backreaction on the background fields, the calculation of the particle production can be trusted only while the energy is smaller than the background energy (dashed line on the plots).  Even more, as far as the particle bath is produced by the oscillations in the $H_0$ direction, it is safe to neglect backreaction only while below the energy in $H_0$, c.f.\ Fig.~\ref{bgPlots}.  However, the latter statement is not too constraining, as far as the effective tachyonic production happens only when the transfer of energy into $H_0$ is effective.

As far as the produced gauge bosons are not relativistic, a mechanism to transfer energy into light particles (radiation) is required.
It is provided by the decays of the produced gauge bosons, which are short lived, and decay within one oscillation of $\Phi_0$.
Indeed, the time-averaged decay width \cite{Tanabashi:2018oca} of longitudinal $W$ bosons
\begin{equation} \label{gammaW}
    \left<\Gamma_W\right>_T \simeq 0.8\,\alpha_W\left<m_W\right>_T
    \sim \omega_{\Phi_0},
\end{equation}
does not allow to accumulate them over more than a single scalaron oscillation. Here, we have used $\alpha_W \approx 0.025$. Meanwhile, we estimate the time-averaged Higgs decay width \cite{Tanabashi:2018oca} to be 
\begin{equation}
    \left<\Gamma_H\right>_T \simeq 0.1\,y_b^2\left< m_H\right>_T
    \ll \omega_{\Phi_0},
\end{equation} 
with Yukawa of $b$-quark $y_b\sim 0.02$, so we need not worry about Higgs decays over preheating timescales.

We therefore see that the closer one approaches the special values of $\beta$ which allow the background to be significantly drained, the faster this draining occurs. What's more, this process is unambiguously driven by the tachyonic enhancement of the inhomogeneous modes.

\begin{figure*}[htb]
	\centering
	\includegraphics[width=0.48\textwidth]{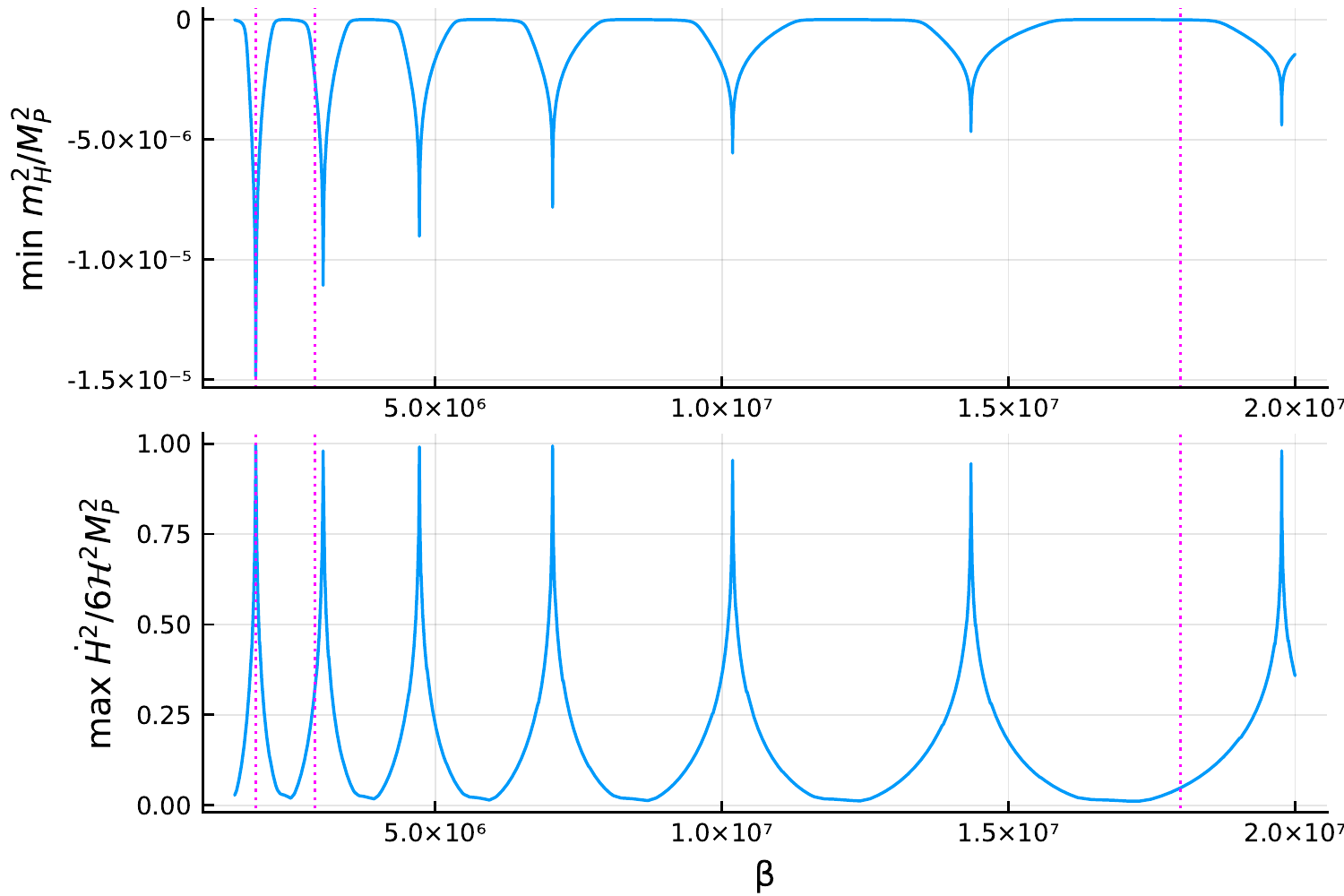}
	\caption{The tachyonic mass after the first zero crossing (top pane) and the part of the energy transferred to the motion in the $H_0$ direction (bottom pane) for a range of $\beta$ close to the Higgs-like inflationary regime. Dashed vertical lines mark the three reference values of $\beta$, $(\beta_1, \beta_2, \beta_3)$, used in the other plots.}
	\label{peaks}
\end{figure*}

One must now consider the generality of this highly-tuned and efficient reheating. Typically, the homogeneous fields will not reflect exactly along the line $H_0=0$; the system will either overshoot or undershoot and end up with a significantly smaller energy in $H_0$, and, therefore, in radiation. The depth of the tachyonic mass dip of the Higgs mode between the first and second scalaron zero crossing, as a function of $\beta$, is shown in Fig.~\ref{peaks}.
Also plotted is the maximum kinetic energy in the background Higgs direction over the positive half oscillation of the scalaron -- the equivalence between the two quantities is clear. Each next peak on Fig.~\ref{peaks} corresponds to one more oscillation of field $H_0$ within the negative half-oscillation of $\Phi_0$, c.f.\ Fig.~\ref{bgZoom}.

Overall, we see that if the zero crossing happened near the bifurcation, energy is efficiently transferred from the background oscillations to the inhomogeneous modes.  Longitudinal gauge bosons are produced more effectively (recall there are three longitudinal components for each of $W^+$, $W^-$ and $Z$-bosons) and rapidly decay to light relativistic particles, leading to full preheating.  Away from the bifurcation, neither is the tachyonic production effective, nor is the amount of energy in the Higgs-like oscillation sufficient for preheating. In this case, one must consider the dynamics of subsequent scalaron oscillations.

Between any two scalaron zero crossings, the non-relativistic longitudinal weak bosons will decay into relativistic products, and the amplitude of the background Higgs mode will fall to nearly zero. Therefore, it can be expected that the system partially ``resets'' its homogeneous background to the smooth motion along the potential valley \eqref{hmin} upon each scalaron crossing -- up to the production of long-lived inhomogeneity in the Higgs mode. Of course, the precise extent to which this resetting occurs cannot be known without  accounting for backreaction. However, the exponential nature of the particle production suggests that resetting should occur near maximally, and one therefore expects a similar profile of peaks to Fig.~\ref{peaks} at subsequent zero crossings. The specifics of the profile at a given scalaron zero crossing depend on the phase of background Higgs oscillations at the moment of  zero crossing. Given that the background Higgs mode oscillates much faster than the scalaron, this phase is extremely sensitive to modifications of the scalaron period.

Without accounting for backreaction, this phase cannot be directly determined, as can be seen from the third term of \eqref{quartic-potential}. The frequency of scalaron oscillation receives a negative contribution from the time-average of $\left(H_0 - H_{min}\right)^2$\footnote{Again, we take the positive $H_{min}$ branch for notational simplicity}. When the energy in the homogeneous Higgs mode is comparable with the total background energy, this contribution is of order unity. Therefore, between the two limits of zero drain and total drain of the homogeneous Higgs mode between scalaron zero crossings, the phase of the homogeneous Higgs at the moment of crossing will vary stochastically. Within the framework of this paper, we will therefore assume this phase to be random.

The above estimations are supported by our numerical studies. Having artificially drained the background Higgs direction in varying amounts between scalaron zero crossings, we have found the profile of peaks in Fig.~\ref{peaks} to be general. The main effect of draining the background Higgs mode is to translate the peak centres to different values of $\beta$ at any particular scalaron crossing, in a manner that appears to be stochastic. Therefore, while one cannot put complete faith in any specific equivalent to Fig.~\ref{peaks} after the first scalaron crossing, one expects the profile of this plot to repeat, with the positions of the peak centres varying randomly between particle production events. Within these estimates for backreaction, we conclude that our specially-tuned situation for tachyonic preheating becomes general if one waits a few scalaron oscillations for it to be realized. This statement is more robust in the Higgs-like regime, because the tachyonic peaks are more abundant in this region of parameter space. In any case, one expects the preheating timescale in the Higgs-like limit to be shorter than the Hubble time, and therefore instantaneous from a cosmological perspective.


\section{Conclusions}
\label{Sec:Conclusion}

We have found that the dynamics of preheating in Higgs inflation regularized by the additional $R^2$ term are quite involved.  Without the regularizing term the system had a spike like feature at zero crossings of the field in the longitudinal gauge boson mass.  In the regularized version, this feature is also present, and corresponds to the motion at small values of the Higgs field and negative values of the scalaron field, however it does not lead to significant particle production.  We demonstrated that immediately after the reflection from the scalaron potential and return to the positive values of the scalaron, the system \emph{may} enter a strongly tachyonic regime, leading to near-instant preheating.  Though this happens at the first zero crossing only for model parameters in finely tuned regions, it is expected that this special scenario is realized in any case after several zero crossings. This conclusion is most robust in the Higgs-like limit, for which we can affirm that preheating is cosmologically instantaneous. With this remark we fix completely the post-inflationary evolution of the model, with reheating temperature $T_{\mathrm{reh}}\simeq 10^{15}$\,GeV and hence the number of e-foldings $N_e$ corresponding to the scale of matter perturbations adopted as the Planck prior. Namely, from eqs.~\eqref{R2-Higgs-predictions} we obtain (c.f.~\cite{Bezrukov:2011gp,Gorbunov:2012ns})
\[
  N_e= 59,\quad n_s=0.97,\quad r= 0.0034.
\]

The relevant dynamics in the system are due to the tachyonic masses appearing in the longitudinal gauge bosons (or, Goldstone bosons in the simpler model without gauge symmetry), and, to probably a lesser extent due to tachyonic mass in the Higgs boson itself.  Further study with full account for backreaction on the background evolution of the fields is required for analysis at any values of the model parameters, allowing one to consistently deal with the system with equal energy distribution between the homogeneous Higgs motion and inhomogeneous modes (weak gauge bosons and Higgs particles).

The work of FB was supported in part by the STFC research grant ST/P000800/1.
The part of the work devoted to production of the weak gauge bosons was supported by Russian Science Foundation grant 19-12-00393.

\providecommand{\href}[2]{#2}\begingroup\raggedright\endgroup

\end{document}